\documentclass[11pt,a4paper]{article}
\usepackage{longtable}
\usepackage{amsmath}
\usepackage{graphicx}
\usepackage{bm}% bold math
\usepackage{footmisc}

\newcommand*{\La}{\cal{L}}
\newcommand*{\no}{\noindent}
\newcommand*{\bea}{\begin{eqnarray}}
\newcommand*{\eea}{\end{eqnarray}}
\newcommand*{\be}{\begin{equation}}
\newcommand*{\ee}{\end{equation}}
\newcommand*{\pd}{\partial}
\newcommand*{\pdm}{\pd_{\mu}}
\newcommand*{\pdn}{\pd_{\nu}}

\newcommand*{\pref}[1]{(\ref{#1})}

\newcommand*{\mn}{{\mu\nu}}

\newcommand*{\nn}{\nonumber}

\newcommand*{\tr}{\mathrm{tr}}

\title{Bound-state/elementary-particle duality in the Higgs sector and the case for an excited 'Higgs' within the standard model}

\author{Axel Maas\\
Institute for Theoretical Physics, Friedrich-Schiller-University Jena,\\
Max-Wien-Platz 1, D-07743 Jena, Germany}

\begin{document}

\maketitle

\begin{abstract}
Though being weakly interacting, QED can support bound states. In principle, this can be expected for the weak interactions in the Higgs sector as well. In fact, it has been argued long ago that there should be a duality between bound states and the elementary particles in this sector, at least in leading order in an expansion in the Higgs condensate. Whether this remains true beyond the leading order is investigated using lattice simulations, and support is found. This provides a natural interpretation of peaks in cross sections as bound states. Unambiguously, this would imply the existence of (possibly very broad) resonances of Higgs and $W$ and $Z$ bound states within the standard model.
\end{abstract}

\section{Introduction}

QED is a remarkable theory. Even though its interactions are so weak that perturbation theory is highly successful in describing it, it harbors a plethora of non-perturbative effects. Foremost among them is the existence of highly complicated bound states, starting with positronium, and encompassing all atoms, molecules, and solids. It is thus a natural question to pose whether the weak interaction can also provide bound states. In fact, from a fundamental point of view, it has been argued that only such bound states can become the gauge-invariant, physical degrees of freedom of such a theory \cite{Frohlich:1981yi,Frohlich:1980gj,'tHooft:1979bj,Banks:1979fi}. Furthermore, almost local gauge-invariant elementary states, as can be constructed in Abelian QED, likely cannot be constructed for the present non-Abelian gauge theory \cite{Haag:1992hx}.

A bit less formal, but still rigorous, argument for the presence of bound states and the absence of the elementary states from the physical spectrum is the Fradkin-Shenker theorem \cite{Fradkin:1978dv}. It states that with a lattice cutoff the partition function of the pure Higgs sector, i.\ e.\ containing only the SU(2) gauge bosons and the fundamental Higgs doublet, is analytic for all values of the (bare) coupling constants. This has two consequences. First, there is no physical distinction between a would-be confinement and a would-be Higgs phase \cite{Caudy:2007sf}. Thus, the asymptotic spectrum is qualitatively the same, and thus free of the elementary particles. Only bound states can be detected asymptotically. Second, these asymptotic bound states are thus qualitatively independent of the strength of the Higgs coupling.

This is a genuine field-theoretical feature, and thus any kind of quantum-mechanical approach will not show this behavior. However, because of the possible triviality of this theory, such a result may depend on the type of regulator. Pending a proof to the contrary or an ultraviolet completion of the theory, it will be assumed here that the issue of triviality has no impact to the following.

Indeed, such bound states already have been observed long ago in lattice calculations see e.\ g.\ \cite{Langguth:1985dr,Jersak:1985nf,Philipsen:1996af}. There, the pure Higgs sector was simulated, i.\ e.\ a theory described by the Lagrangian
\bea
{\La}&=&-\frac{1}{4}W_\mn^aW^\mn_a+(D_\mu\phi)^+D^\mu\phi-\gamma(\phi\phi^+)^2-\frac{m^2}{2}\phi\phi^+\label{action}\\
W_\mn^a&=&\pdm W_\nu^a-\pdn W_\mu^a-g f^{abc} W_\mu^b W_\nu^c\nn\\
D_\mu^{ij}&=&\pd_\mu\delta^{ij}-igW_\mu^a\tau^{ij}_a\nn,
\eea
\no with the fundamental complex scalar $\phi$ coupled to su(2) valued gauge fields $W_\mu$ with the field strength tensor $W^a_\mn$, the covariant derivative $D_\mu$, and the coupling constants $g$, $\gamma$, and $m$.

Such gauge-invariant bound states include $0^{++}$ Higgs\-onium states \cite{Langguth:1985dr,Jersak:1985nf,Philipsen:1996af}, created by the operator
\be
{\cal O}^{0++}=\phi_i^+(x)\phi^i(x)\label{higgsonium}
\ee
\no with $\phi_i$ the Higgs field, the $0^{++}$ $W$-ball created by the operator $W_\mn^a(x) W^\mn_a(x)$ from the field strength tensor $W_\mn^a$ of the $W_\mu^a$ field \cite{Philipsen:1996af}, and the $1^{--}$ state created by a more complicated operator, written in continuum \cite{Dosch:1983hr,Dosch:1984ec}
\bea
{\cal O}^{1--}_{a\mu}&=&\tr\tau^a\varphi^+ D_\mu\varphi\label{w}
\eea
\no or lattice notation as \cite{Langguth:1985dr,Jersak:1985nf,Philipsen:1996af}
\be
{\cal O}^{1--}_{a\mu}=\tr \tau_a \varphi^+(x) \exp(i\tau_bW^b_\mu(x))\varphi(x+e_\mu).\label{wl}
\ee
\no Herein, the Higgs field has been decomposed in its length $|\phi|=\rho$ and its direction, which is encoded in $\varphi$ as an SU(2) matrix \cite{Langguth:1985dr,Jersak:1985nf}, and $D_\mu$ is the covariant derivative. The $\tau^a$ are the Pauli matrices, and $e_\mu$ is a unit vector in the direction $\mu$. It should be noted that $a$ is not a gauge index, but refers to the global flavor symmetry of the pure Higgs sector \cite{Shifman:2012zz}. Results from the pure Higgs sector are not quantitatively reliable for comparison to the experiment, as they neglect e.\ g.\ fermions and QED\footnote{Which implies that the $W$ and $Z$ are degenerate.}. This implies, of course, that especially the widths of these states cannot be translated to the standard model, since at least the lightest bound state is stable.

The structure of such bound states is, due to the Higgs effect, profoundly different from those of QCD or QED. This will be discussed in section \ref{sdual}, and has been predicted already more than thirty years ago. Here, these predictions will be tested and supported using lattice gauge theory, see section \ref{slattice}. The consequences of this structure will be discussed in section \ref{sconsequences}, and briefly summarized in section \ref{sconclusion}. The technical details of the lattice simulations presented are almost identical to the ones of \cite{Cucchieri:2006tf,Maas:2010nc,Philipsen:1996af}, and therefore deferred to the appendix \ref{sasd}.

\section{Bound-state-elementary-state duality}\label{sdual}

The natural question is then, whether a relation between the masses of the elementary states and of the bound states exist, like in a constituent quark model of QCD. A simple translation of the idea of the quark model turns out to be likely not applicable \cite{Maas:2011jy}\footnote{Note that there is a slight quantitative error in the preliminary data on the Schwinger functions reported in \cite{Maas:2011jy}, which does not affect the conclusions.}. However, a relation has been proposed long ago \cite{Frohlich:1981yi,Frohlich:1980gj} which implies that the bound state masses and the elementary particle masses are related.

For the following, it is necessary to select an adequate gauge, e.\ g.\ 't Hooft gauge \cite{Bohm:2001yx} or an aligned Landau gauge \cite{Maas:2012ct}. The Higgs field in the Higgs phase of this gauge has a non-vanishing expectation value of size $v$ with arbitrary, but fixed, direction $n$ in isospin space. Expand now the Higgs field $\phi$ in its fluctuation around this expectation value, $\phi^i=v n^i+\eta^i$, with the fluctuation field $\eta$. The Higgsonium correlator, based on the composite operator \pref{higgsonium}, then becomes\setcounter{footnote}{3}\footnotetext{\label{ope}See \cite{Dosch:1983hr,Dosch:1984ec} for similar considerations in the context of an operator product expansion approach in the Abbott-Farhi model \cite{Abbott:1981re}.}\footref{ope}
\bea
&&\langle\phi_i^+(x)\phi^i(x)\phi_j^+(y)\phi^j(y)\rangle\nn\\
&\approx& v^4+4v^2(c+\langle\eta^+_i(x) n^i n_j^+\eta_j(y)\rangle)+{\cal O}(\eta^3)\label{correl},
\eea
\no with $c$ some irrelevant constant. Thus, to this order, the correlation functions of the bound state and the elementary particle are similar. From this follows that any pole in the bound state on the left-hand side will be reflected by a pole of the elementary particle on the right-hand side. This is a duality between the bound state and the elementary state. Of course, since the expansion in $\eta$ is an expansion in the quantum fields, this relation can only be expected to hold exactly at tree-level. The same can be repeated for \pref{w} and \pref{wl} \cite{Frohlich:1981yi,Frohlich:1980gj}. However, since the $W$ field has no expectation value, to lowest order the angular part of the Higgs field $\varphi$ is just replaced with an SU(2) representation of $n$ in \pref{w} and \pref{wl}, and thus the $1^{--}$ correlator becomes the one of the $W$ boson\footref{ope},
\bea
&&\langle(\tau^a\varphi D_\mu\varphi)(x)(\tau^a\varphi D_\mu\varphi)(y)\rangle\nn\\
&\approx& \tilde{c}\tr(\tau^a\tilde{n}\tau^b\tilde{n}\tau^a\tilde{n}\tau^c\tilde{n})\langle W^b_\mu W^c_\mu\rangle+{\cal O}(\eta W)\label{correl2},
\eea
\no where $\tilde{c}$ is a new constant, and $\tilde{n}$ is the SU(2) representation of the constant (four-dimensional) unit vector $n$. Note that to leading order in the Higgs vacuum expectation value $\pdm \phi\approx\pdm v=0$ and $\pdm\varphi\approx\pdm\tilde{n}=0$. Thus again the pole masses of the bound state and the elementary state should coincide to leading order. A similar argument cannot be made for the $W$-ball, as the $W$ field cannot be expanded likewise.

It should be noted that if \pref{correl} is averaged over all possible orientations of $n_i$, i.\ e.\ making a non-aligned global gauge choice \cite{Maas:2012ct}, the relation turns into
\be
\langle\phi_i^+(x)\phi^i(x)\phi_j^+(y)\phi^j(y)\rangle\approx v^4+v^2(c'+\langle\phi^+(x)\phi(y)\rangle)+{\cal O}(\eta)\label{correl3}.
\ee
\no This implies that the relation is independent of the choice of the global gauge condition.

\section{Test using lattice calculations}\label{slattice}

Beyond this leading order, the dependence of the Higgs pole mass on the renormalization scheme for the Higgs \cite{Bohm:2001yx} immediately implies that a relation like \pref{correl} can no longer hold. However, it is, of course, possible to use a renormalization scheme in which the pole mass of the renormalized Higgs propagator is the pole mass of the Higgsonium. This is different for the $W$ boson, since no independent mass renormalization is possible \cite{Bohm:2001yx}.

The question is thus, what happens beyond tree-level to the relations \pref{correl} and \pref{correl2}. To test this, standard lattice calculations can be used to determine the renormalized correlation functions of the bound bound states \cite{Gattringer:2010zz} and the elementary states \cite{Maas:2010nc,Maas:2011se}, and from them the corresponding masses \cite{Maas:2011se,Gattringer:2010zz}. This is done here, see appendix \ref{sasd} for the technical details. Since the relations are only expected to hold when the quantum fluctuations are small for the Higgs, a corresponding system has to be simulated. In the present case, the average fluctuations are found to be of the order of 1\%, if one determines $\langle|\eta|\rangle/\langle|\phi|\rangle$, though this is a gauge-dependent statement, and the associated error is of similar size. Of course, it cannot be expected that the approximate relation \pref{correl} holds once $\eta/v$ is no longer small, i.\ e., in the presence of strong (Higgs-self-)interactions. For a tree-level Higgs mass of 125 GeV, the relation should still hold true.

\begin{figure}
 \includegraphics[width=0.5\linewidth]{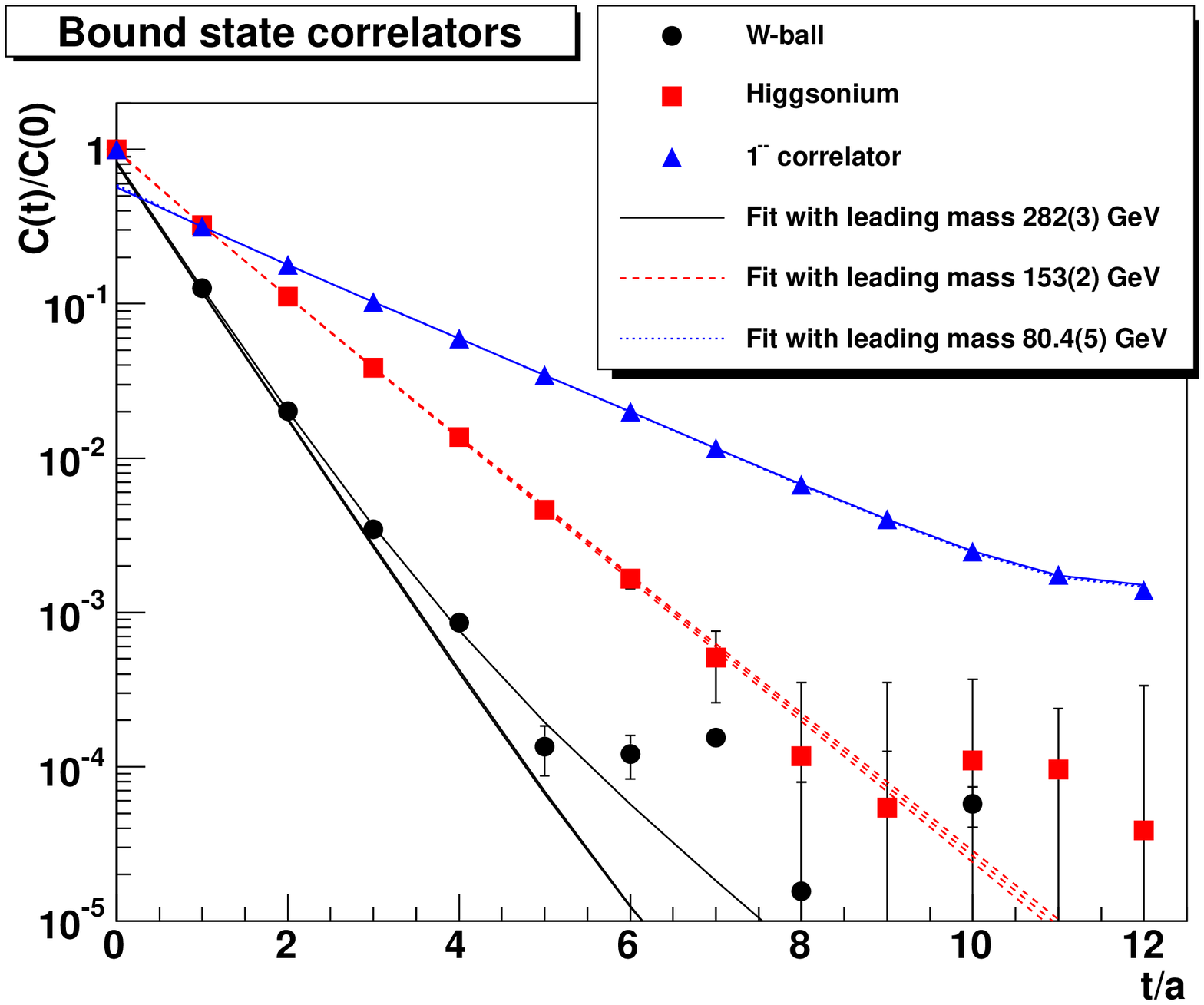}\includegraphics[width=0.5\linewidth]{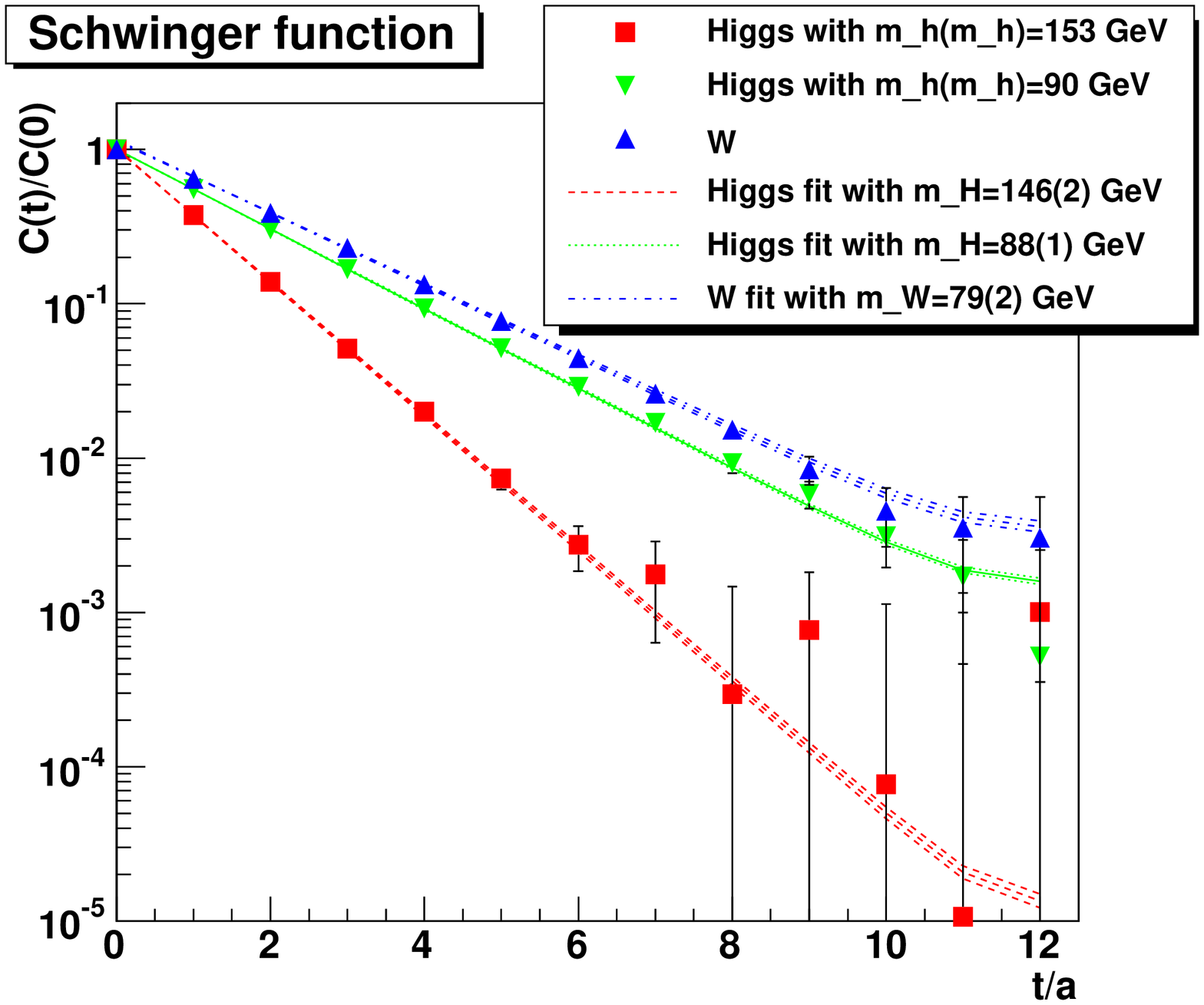}
 \caption{\label{cor}Results from the lattice with a lattice cutoff $a$. The left panel shows the correlators for the bound states. The significance of the also shown $W$-ball correlator will be discussed in section \ref{sconsequences}. The right panel shows the Schwinger correlation function of the elementary states \cite{Maas:2011se}, which have been obtained from the renormalized propagators. See appendix \ref{sasd} for the technical details of obtaining these correlation functions and their fits.}
\end{figure}

The resulting correlation functions for the elementary particles, the $W$ and the Higgs, and for the physical bound states in the $0^{++}$ and $1^{--}$ channels are shown in figure \ref{cor}. Always shown are also fits where the fit form is the one expected for a massive particle \cite{Gattringer:2010zz,Montvay:1994cy}. Note that statistical fluctuations are large and one should be wary of systematic errors \cite{Maas:2011se,Gattringer:2010zz}. These will be explored in the future \cite{Maas:unpublished}, as the primary aim here is just a proof-of-principle. Some preliminary investigations have shown no significant influence of lattice size or discretization.

Since the $W$ is more easy to interpret, it will be investigated first. The $W$ correlator exhibits a mass which is, within errors, identical to the one in the $1^{--}$ channel, confirming \pref{correl2}. It is thus used to set also the scale to determine the masses in physical units.

The situation for the elementary Higgs is somewhat more subtle. Since its mass depends on the renormalization scheme, the associated pole mass is also shifted by a scheme change, as is visible on the right hand side of figure \ref{cor}. The mass of the associated $0^{++}$ bound state, on the other hand, is well-defined, and found to be about 153 GeV, just below the 2 $1^{--}$ bound state threshold. As expected, renormalization effects thus spoil the relation \pref{correl} beyond tree-level, though there exists a scheme in which it still holds.

Thus, the proposed approximate equality \cite{Frohlich:1981yi,Frohlich:1980gj} between the bound state masses and the elementary particles pole position \pref{correl2} and  \pref{correl}, at least in an adequate renormalization scheme, is supported beyond tree-level.

If, as in QCD, the $0^{++}$ and $1^{--}$ bound states should be interpreted in a loose sense as made up at tree-level from 2 Higgs and 2 Higgs and 1 $W$, respectively, this would imply that the mass defect is of the order of the mass of the constituents. Thus, these states are deeply bound, relativistic states, invalidating any attempt to use quantum mechanical estimates to describe them, like was done to exclude such bound states in the past, see e.\ g.\ \cite{Grifols:1991gw}.

\section{Consequences}\label{sconsequences}

This has quite interesting, though speculative, consequences\footnote{Note that a similar idea as the following has been put forward in \cite{Arbuzov:2012kb}, but there a Higgs candidate is interpreted as what here is called the $W$-ball. This would be an option, if the Higgsonium were heavier, which at least for a number of parameter settings is not the case \cite{Maas:2011jy,Maas:unpublished}. It also leaves out the question of the nature of the $W$ itself. Furthermore, such states may mix with heavy quarkonia resonances, as proposed to be observable e.\ g.\ in \cite{Moffat:2012ha}, or other electroweak bound states \cite{Ellis:2012io}.}. One is led to these, when one starts asking, what it actually is that is measured in an experiment. Consider the Higgs case, where the problem is more obvious. The Higgs pole mass is, as seen in figure \ref{cor}, evidently dependent on the renormalization scheme, and thus also on the renormalization scale. When making an experiment, what is observed is a resonance in the cross section, ideally in a certain channel with given quantum numbers. This resonance is not dependent on the gauge, renormalization scale, or renormalization scheme. At tree-level, the resonance is directly associated with the corresponding elementary particle, say the Higgs. Beyond tree-level, the renormalization-scheme dependence spoils this 1-to-1 relation, and the scheme dependence is canceled only in the final cross section. Though, even if the Higgs has the arbitrary pole mass $m_h$, chosen by our choice of scheme, the actually observed mass of the resonance in the cross section $m_\sigma$ is not the same, in general, except in a particular scheme. Since the scheme choice cannot entail physics, what is the actual physical resonance observed in the experiment? The only logical conclusion is that this is actually a bound state mass with the same quantum numbers.

\begin{figure}
\begin{center}
 \includegraphics[width=0.5\linewidth]{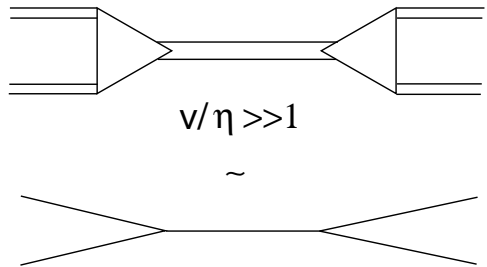}
\end{center}
 \caption{\label{corscatt}The scattering of two gauge-invariant initial bound states to final gauge-invariant bound states of e.\ g.\ two fermions can proceed via the formation of an intermediate bound state. This will manifest itself as a resonance. However, when expanding in leading order for small quantum fluctuations of the Higgs field, the process becomes equivalent to the tree-level perturbative one, manifesting that the resonance has at tree-level the tree-level Higgs mass. Double lines are bound states of anything with a Higgs, the triangles correspond to the Bethe-Salpeter amplitudes, and the thin lines to the tree-level process.}
\end{figure}

Take a $f\bar f\to \tilde{f}\bar{\tilde f}$ scattering, with some fermions $f$ and $\tilde f$ in the initial and final states. To obtain gauge-invariant initial and final states, these have to be interpreted as Higgs-fermion bound states once more, where a similar duality holds between the fermion pole masses and the bound state masses \cite{Frohlich:1981yi,Frohlich:1980gj}, though this is of no direct relevance here. Thus, the scattering has to be regarded as a bound-state scattering, where in an intermediate state a further bound state, e.\ g.\ a Higgsonium, is formed, see schematically figure \ref{corscatt}. This is very similar to the formation of, e.\ g., a quarkonium as an intermediate state in the corresponding QCD process.

In case of the $W$ boson, the situation is more subtle. In fact, its pole mass may in general not be gauge-invariant, even if it is gauge-parameter-independent in covariant gauges \cite{Nielsen:1975fs}. Thus, even if its pole mass is independent of the renormalization scheme and scale, it is not necessarily physical. Nonetheless, a similar argumentation can be made that the $W$ boson identified in the scattering cross section is just the corresponding $1^{--}$ bound state \pref{w}.

A very simple question for this scenario is, why the cross section is rather well described by perturbation theory, even if the formation of intermediate bound states is a genuinely non-perturbative process? At tree level, this is a direct consequence of the relation \pref{correl}. To really answer this question beyond tree level will require to calculate a process like the one shown in the upper part of figure \ref{corscatt}, and then compare its expansion in terms of the Higgs expectation value with the perturbative expression. This will require to make a precise statement how, e.\ g., residual renormalization scale and scheme dependencies in the perturbative calculation, which can hide deviations between both results, have to be handled. It is non-trivial to estimate the size of these effects, and thus how much they can shadow the non-perturbative corrections, see e.\ g.\ \cite{Baglio:2010um}. That is a necessary, but very complicated task \cite{Luscher:1990ux,Arbuzov:2012kb}, not quickly to be solved, even in an approximation where most of the standard model is neglected. But it can be expected that the non-perturbative contributions should be suppressed by at least one power of either the weak fine-structure constant or the four-Higgs coupling, making it at most a percent effect in nature. Nonetheless, it cannot be expected that a close similarity will hold if there are strong non-perturbative contributions in the Higgs sector, and thus it appears likely that the actual Higgs(onium) is light indeed.

The ultimate consequence of this discussion of the bound-state-elementary-state duality is then the role of the $W$-ball. It is a heavier, excited state, at least for the presently employed lattice parameters. Thus, it is likely unstable, and will decay quickly. However, if it exists sufficiently long, then it can be produced in the same way as the Higgsonium. An estimate of its width could be obtained using lattice methods along the lines of \cite{Luscher:1990ux,Luscher:1991cf}. Hence, similar again to the quarkonium case, it can be expected that excited states will turn up as additional resonances. If the lattice results here give any impression of the actual scales, the $W$-ball will be much heavier than the Higgsonium, about twice its mass. Thus, another resonance may be expected to turn up at the LHC at a mass of, very speculative, 200-300 GeV assuming a 125 GeV Higgs(onium), though much broader than the Higgs. Similarly, an excited $W$ can exist. The fits in figure \ref{cor} indicate a mass of more than 280 GeV, though strong lattice artifacts can be expected for such a heavy particle. Because of the resulting large phase space and the open decay channels, these excitations are likely rather unstable, though in the ungauged Higgs-Yukawa model even for large phase space small decay widths have been observed \cite{Gerhold:2011mx}. However, the experimental searches at the LHC in both the Higgs channel \cite{ATLAS:2012ae,CMS-PAS-HIG-12-008} and for an excited weak gauge boson \cite{Aad:2011yg,Chatrchyan:2012qk} have not yet shown any evidence for any kind of particles. If the masses given here are anywhere reasonable, this implies that either their production cross section is small or their decay width is indeed large. In any case, this implies that a search at the LHC will be challenging. On-resonance production at the ILC, however, maybe a more promising possibility. Furthermore, there may be further bound states with more exotic quantum numbers \cite{Philipsen:1996af}. Though there production will again be a higher order process, they may be a further signature, a possibility currently under investigations \cite{Maas:unpublished}.

This type of resonances, essentially an 'excited Higgs' and an 'excited W', constitute a standard model background. Given that many beyond-the-standard-model scenarios introduce new particles in both channels \cite{Morrissey:2009tf,Andersen:2011yj}, a full understanding of the possible existence and properties of heavy standard model resonances is mandatory, to exclude this possibility before new physics can be unambiguously claimed in case of an observation. It is a challenging task to determine the properties and existence of such excited states in the standard model, but it is a necessary task.

Furthermore, the duality provides a new take on the hierarchy problem. If the standard model is non-trivial, the bound state masses are independent of the renormalization scale and the renormalization scheme. They are not affected in the same way as the elementary states by the hierarchy problem, which may therefore be alleviated for observable particles. But this requires a better understanding of the triviality problem and the spectrum of the bound states near the continuum.

\section{Conclusions}\label{sconclusion}

Summarizing, the structure of quantum gauge theories dictates that asymptotic states, even in the presence of a Higgs effect, have to be gauge-invariant. Based on lattice arguments \cite{Fradkin:1978dv,Caudy:2007sf}, the asymptotic states are likely bound states of Higgs and $W$ particles. For such bound states and for sufficiently light Higgs particles an interesting relation holds between their mass and the tree-level elementary particle mass  \cite{Frohlich:1981yi,Frohlich:1980gj}, implying their equality. Here, it has been shown, using lattice gauge theory, that this equality holds almost even for full correlation functions. This opens up a possibility how to interpret the observed particles in experiments in terms of gauge-invariant, physical states, instead of the gauge-dependent elementary particles. However, it will require much better systematic investigations to fully understand this, and to make reliable predictions for experiments.\\

\no{\bf Acknowledgments}

I am grateful to J\"org J\"ackel, Sebastian J\"ager, and Roman Zwicky for helpful discussions. This work was supported by the DFG under grant number MA 3935/5-1. Simulations were performed on the HPC cluster at the University of Jena.

\appendix

\section{Simulation details}\label{sasd}

\subsection{Simulations}

The lattice simulations presented are based on an unimproved lattice version of the action \pref{action}, given by \cite{Montvay:1994cy},
\bea
S&=&\beta\sum_x\Big(1-\frac{1}{2}\sum_{\mu<\nu}\Re\tr U_\mn(x)+\phi^+(x)\phi(x)+\lambda\left(\phi(x)^+\phi(x)-1\right)^2\label{laction}\\
&&-\kappa\sum_\mu\Big(\phi(x)^+U_\mu(x)\phi(x+e_\mu)+\phi(x+e_\mu)^+U_\mu(x)^+\phi(x)\Big)\nn\\
U_\mn(x)&=&U_\mu(x)U_\nu(x+e_\mu)U_\mu(x+e_\nu)^+U_\nu(x)^+\nn\\
W_\mu&=&\frac{1}{2agi}(U_\mu(x)-U_\mu(x)^+)+{\cal O}(a^2)\label{wdef}\\
\beta&=&\frac{4}{g^2}\nn\\
a^2m_0^2&=&\frac{(1-2\lambda)}{\kappa}-8\nn.
\eea
\no In this expression $a$ is the lattice spacing, $W_\mu$ the gauge boson field, $U_\mu$ the corresponding link variable $\exp(iga W_\mu)$, $\phi$ the Higgs field, $g$ the bare gauge coupling, $\lambda/\kappa^2$ is the bare self-interaction coupling of the Higgs, $m_0$ the bare mass of the Higgs, the sums are over the lattice points $x$, and $e_\mu$ are unit vectors on the lattice in the direction $\mu$. The simulations have been performed with bare gauge coupling $g(a)=1.32$, bare 4-Higgs coupling $\gamma(a)=9.77$, and bare Higgs mass $m(a)^2=-(487$ GeV$)^2$, i.\ e.\ $\beta=2.3$, $\kappa=0.32$, and $\lambda=1$ in the lattice notation \pref{laction} on a $N^4=24^4$ lattice. Lattice simulations with smaller statistics on $20^4$ and $28^4$ lattices did not show, within statistical errors, any deviations for the final results.

The generation of configurations has been performed as in \cite{Maas:2010nc}, using a combination of one heat-bath and five-overrelaxation sweeps for the gauge fields according to \cite{Cucchieri:2006tf}, and between each of these 6 sweeps of the gauge fields one Metropolis sweep for the Higgs field using a Gaussian proposal with a self-tuned width set to 50\% acceptance probability. These updates have been performed lexicographical. These 12 sweeps together constitute a single update for the field configuration. The auto-correlation time of the plaquette is of the order of 1 such update. Thus, 24 such updates separate a measurement of a gauge-invariant observable, to reduce the auto-correlation time. Because of the strong statistical fluctuations of the correlators evaluated here, these have not been used to determine the auto-correlation time. For the thermalization, 1080 such updates have been performed. In total, 1249434 independent configurations have been created in in total 1192 runs.

All errors have been calculated using bootstrap with 1000 re-samplings and give a 67.5\% interval, i.\ e.\ approximately 1$\sigma$ interval.

\subsection{Gauge-invariant correlators}

In the $0^{++}$ channel two correlation functions have been determined. One is the Higgsonium operator \pref{higgsonium}, the other the $0^{++}$ $W$-ball state with the plaquette as the lattice discretization of $W_\mn^a W_\mn^a$ \cite{DeGrand:2006zz}. Since the operators are very noisy, they have been five-time APE smeared, i.\ e.\ the operators have been measured using the smeared links and Higgs fields \cite{Philipsen:1996af}
\bea
U_\mu(x)^{(n)}&=&\frac{1}{\sqrt{\det V_\mu(x)^{(n)}}}V_\mu(x)^{(n)}\nn\\
V_\mu^{(n)}&=&\alpha U_\mu(x)^{(n-1)}+\frac{1-\alpha}{2(d-1)}\times\sum_{\nu\neq\mu}\left(U^{(n-1)}_\nu(x+e_\mu)U^{(n-1)+}_\mu(x+e_\nu)U^{(n-1)+}_\nu(x)\right.\nn\\
&&\left.+U_\nu^{(n-1)+}(x+e_\mu-e_\nu)U_\mu^{(n-1)+}(x-e_\nu)U_\nu^{(n-1)}(x-e_\nu)\right)\nn\\
\phi^{(n)}&=&\frac{1}{1+2(d-1)}\left(\phi^{(n-1)}+\sum_\mu(U^{(n-1)}_\mu(x)\phi^{(n-1)}(x+e_\mu)+U^{(n-1)}_\mu(x-e_\mu)\phi^{(n-1)}(x-e_\mu))\right)\nn,
\eea
\no with $\alpha=0.55$ and $d=4$.

\begin{figure}
\begin{center}
 \includegraphics[width=0.5\linewidth]{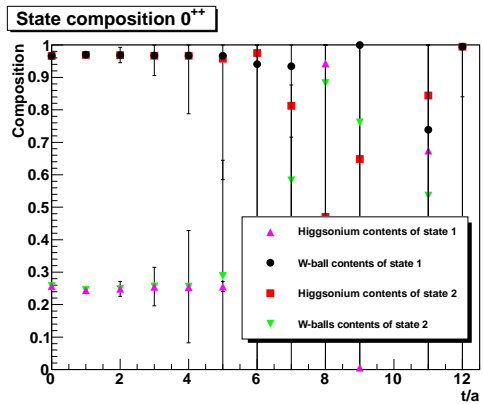}
\end{center}
 \caption{\label{composition}The eigenvector components of the correlation matrix of the $0^{++}$ states as a function of the lattice time, for details see text.}
\end{figure}

To disentangle the ground state and the first excited state the full correlation matrix of both operators has been determined and the corresponding eigenvalues and eigenfunctions have been determined \cite{Gattringer:2010zz}. The eigenvectors, shown in figure \ref{composition}, show little mixing between the Higgsonium and the $W$-ball states, with the Higgsonium being the ground state and the $W$-ball the first excited state\footnote{It cannot be excluded that the first excited state is actually a scattering state. Future investigations will have to show this.}. The eigenvalues as a function of lattice time have been shown in figure \ref{cor}.

\begin{table}
\caption{\label{fits}The fit parameters for the different correlation functions. The masses are also given in physical units with a scale set to $a^{-1}=146.7$ GeV, as discussed in the text. Errors were determined by shifting the correlators within statistical errors upwards and downwards. In case of the gauge-dependent correlators a reliable fitting of subleading contributions was notpossible.}
\vspace{1mm}
\begin{tabular}{|c|c|c|c|c|c|c|}
\hline
Object & $A$ & $am$ & $m$ [GeV] & $B$ & $an$ & $n$ [GeV] \cr
\hline
Higgsonium & $1.54(5)\times 10^{5}$ & $1.04(1)$ & $153(2)$ & $0.14(5)$ & $-0.88(3)$ & $129(5)$ \cr
\hline
$W$-ball & $3.52(1)\times 10^{3}$ & $1.92(2)$ & $282(3)$ & $0.02(2)$ & $0.88(3)$ & $129(5)$ \cr
\hline
$1^{--}$ state & $7.1(2)\times 10^{3}$ & $0.548(3)$ & $80.4(5)$ & $0.0001(50)$ & $1.8(5)$ & $264(73)$ \cr
\hline
$W$ & $0.0036(4)$ & $0.538(9)$ & $79(2)$ & & & \cr
\hline
$\stackrel{\text{\large{Higgs}}}{\mu=153\text{ GeV}}$ & $1.3(2)\times 10^-5$ & 0.993(9) & 146(2) & & & \cr
\hline
$\stackrel{\text{\large{Higgs}}}{\mu=90\text{ GeV}}$ & $1.59(9)\times 10^{-3}$ & 0.595(5) & 88(1) & & & \cr
\hline
\end{tabular}
\end{table}

The fits in figure \ref{cor} have been made using the fit ansatz \cite{Gattringer:2010zz}
\be
C(t)=A\exp(-amt)(1+B\exp(-a nt))\nn,
\ee
\no where $n=\Delta m$ was set to the difference between the ground state and the first excited state, which is to be expected if the second excited state should be farther away from the first excited state than the first excited state is from the ground state \cite{Gattringer:2010zz}. Fitting alternatively with $n$ an independent parameter did not improve the fit quality noticeable and especially did not change the value of $m$ appreciably within errors. In all fits the point at $t=0$ was excluded, and the largest $t$ included was the last one where the relative error was smaller than 100\% and the correlator was positive including errors. The fit parameters are given in table \ref{fits}. Errors to the fits have been determined using a correlated shift of the data within the given errors upwards or downwards.

For the vector state, the operator \pref{wl} has been used, again with the five times smeared operators. As noted in the text, this operator can be identified with the observed $W$ resonance in experiments. Its mass has therefore been used to set the scale, which thus has an uncertainty of 0.7\%. Since only this single operator has been used, the fit ansatz was \cite{Gattringer:2010zz}.
\be
C(t)=A\cosh\left(am\left(t-\frac{N}{2}\right)\right)+B\cosh\left(an\left(t-\frac{N}{2}\right)\right)\label{stdfit}.
\ee

\subsection{Gauge-dependent correlators}

To obtain the gauge-dependent correlation functions, a subset of 9901 of these configurations have been gauge-fixed. The local part of the gauge has been fixed to Landau gauge, using a self-tuning stochastic overrelaxation algorithm with a quality parameter $e_6$ smaller than $10^{-12}$, see \cite{Cucchieri:2006tf} for details. Since the number of gauge-fixing sweeps shows in general a larger auto-correlation time than the plaquette \cite{Maas:2008ri}, the number of updates between two consecutive measurements of gauge-dependent observables was at least 162. However, because the relation \pref{wdef} only holds for a positive Polyakov loop \cite{Karsch:1994xh}, configurations with negative Polyakov loop in any direction have not been included for gauge-fixed measurements\footnote{Since the center symmetry is broken explicitly by the Higgs field it is not possible to just rotate the field configurations as in Yang-Mills theory. However, the value of the Polyakov loop is so small that only an upper limit can be determined.}. The global part of the gauge was fixed such that $<\phi_i>=0$ \cite{Maas:2012ct}, i.\ e.\ a non-aligned gauge. This was achieved by a random global gauge transformation after the fixing to Landau gauge. Of course, thies implies that the Higgs expectation value was only vanishing on the average, but was sufficiently small for the present purpose, about 0.04\% of the length $\langle|\phi|\rangle$ of the Higgs field.

The $W$ propagator \cite{Maas:2010nc}
\be
D_\mn^{ab}=\langle W_\mu^a W_\nu^b\rangle
\ee
in the non-aligned Landau gauge is transverse with a single dressing function $Z$, which is multiplicatively renormalized with the wave-function renormalization factor $Z_W$
\be
D_\mn^{ab}=\delta^{ab}\left(\delta_\mn-\frac{p_\mu p_\nu}{p^2}\right)\frac{Z_WZ(p^2)}{p^2}\nn.
\ee
\no The renormalization condition is chosen as 
\be
Z_WZ(\mu)=\frac{\mu^2}{\mu^2+m_{W}^2}\nn,
\ee
\no with $\mu=m_W=80$ GeV and $m_W=80$ GeV. The configuration space correlator is then determined using the prescription
\be
\Delta(t)=\frac{1}{a\pi}\frac{1}{N}\sum_{P_0=0}^{N_t-1}\cos\left(\frac{2\pi tP_0}{N_t}\right)\frac{Z(P_0^2)}{P_0^2}\label{schwinger}.
\ee
\no Of course, since the renormalization is purely multiplicative this yields precisely the ordinary correlation function in position space multiplied by $Z_W$. The fit ansatz in momentum space used is again \pref{stdfit}, with the results shown in table \ref{fits}.

This is different for the Higgs propagator, given by
\be
D_H^{ab}=\langle\phi(p)^{a+}\phi(p)^{b}\rangle\nn.
\ee
\no The renormalization of the Higgs propagator requires besides the multiplicative wave-function renormalization also an additive mass renormalization. The renormalized propagator is given by \cite{Bohm:2001yx}
\be
D_H^{ab}(p^2)=\frac{\delta^{ab}}{Z_H(p^2+m^2)+\Pi_H(p^2)+\delta m^2}\nn,
\ee
\no where $\Pi_H$ is its self-energy, and $Z_H$ and $\delta m^2$ are the wave-function and mass renormalization constants, respectively. The two renormalization conditions implemented are \cite{Maas:2010nc}
\bea
D_H^{ab}(\mu^2)&=&\frac{\delta^{ab}}{\mu^2+m_H^2}\nn\\
\left.\frac{\pd D_H^{ab}(p^2)}{\pd |p|}\right|_{|p|=\mu}&=&-\frac{2\mu\delta^{ab}}{(\mu^2+m_H^2)^2}\nn,
\eea
\no with $m_H=\mu$. Selecting $m_H$ thus corresponds to selecting a (mass-dependent) renormalization scheme. It is this renormalized propagator which is transformed to position space using \pref{schwinger} for the two renormalization conditions given in figure \ref{cor} and fitted using \pref{stdfit}, with parameters shown in table \ref{fits}.

\bibliographystyle{bibstyle}
\bibliography{bib}

%%%%%%%%%%%%%%%%%%%%%%%%%%%%%%%%%%%%%%%%%%%%%%%%%%%%%%%%%%%%%%%%%%%%%%%%%%%%%%%%%%%%%%%%%%%%%%%%%%

\end{document}